# The Full Laplace-Beltrami operator on U(N) and SU(N).

Ole L. Trinhammer

*Rungsted Gymnasium, Stadion Allé 14, 2960 Rungsted Kyst, Denmark*

Gestur Olafsson

*Department of Mathematics, Louisiana State University, Baton Rouge, LA 70803, USA*

The Laplacian on the Lie groups U(N) and SU(N) is given in a parametrized edition for practical purposes. The radial part is often seen in work on lattice gauge theory, but here is derived also the off-diagonal part which in SU(3) and U(3) is expressed via the well known Gell-Mann matrices but with a more easily memorized notation. Relations to *I*, *U* and *V* spin are also shown.

MSC-class: 22E70 (Primary) 43A90 (Secondary)

## I. INTRODUCTION

The radial part of the Laplacian on group spaces like the Lie group U(N) is well known in the physics literature[1,2,3]. Related Jacobians for integration on algebras have been derived explicitly in work on counting graphs on closed Riemannian surfaces[4] relevant in lattice gauge theory.

However to our experience the full Laplacian on U(N) and SU(N) is not well known among physicists (see acknowledgments). We know only of Helgason's work[5] where a general expression is given, though not for compact spaces. From this a specific expression could be found by "analytic continuation" arguments which can relate information on the compact symmetric space SO(n)/SO(n-1) = $S^{n-1}$ and the hyperboloid SO(1,n-1)/SO(n-1). Helgason's work, however, is high brow mathematics and it is not easy for a physicist to read off the expression in specific cases. Thus we will give here a derivation based solely on mathematics which should be more familiar to physicists.

We have two standard expressions for the Laplacian, one in which the space is considered as a Riemanian manifold[5,6,7]

$$\Delta = \frac{1}{\sqrt{\|g\|}} \partial_k \sqrt{\|g\|} g^{kl} \partial_l \qquad (1)$$

and one in which the space is considered as a Lie group[8]

$$\Delta = \tilde{Z}_k \tilde{Z}_k = C_2 \qquad (2)$$

where $\|g\|$ is the absolute value of the determinant of a suitable metric in local coordinates with components $g_{kl}$ and $C_2$ is a Casimir operator constructed from left invariant vector fields $\tilde{Z}_k$.

We want to derive an expression

$$\Delta = \sum_{j=1}^{3} \frac{1}{J^2} \frac{\partial}{\partial \theta_j} J^2 \frac{\partial}{\partial \theta_j} - \sum_{\substack{i<j \\ k \neq i,j}}^{3} \frac{L_k^2 + M_k^2}{8 \sin^2 \frac{1}{2}(\theta_i - \theta_j)}$$

which combines the two views into an analogue of the well known spherical coordinate Laplacian[9]

$$\Delta = \frac{1}{r^2} \frac{\partial}{\partial r} r^2 \frac{\partial}{\partial r} - \frac{1}{r^2} \mathbf{L}^2$$

where the angular momentum operator $\mathbf{L} = (L_x, L_y, L_z)$ acts only in the angular space. $J$ is the socalled van der Monde's determinant (equation 25), $\theta_j$ are the "eigenangles" of the group elements and $L_k$ and $M_k$ are the off diagonal generators. $L_k$ commutes as body fixed angular momentum operators and $M_k$ are mixing operators which "close" the algebra (see equations 38, 39).

## II. POLAR DECOMPOSITION

We will reformulate an old idea by Hermann Weyl[10,11] into modern language. The reformulation is essential in order to specify the non-radial part which has been suggested with great intuition by Spenta R. Wadia[12].

First note that any element $v \in U(N)$ can be represented as an $N \times N$ unitary matrix which is conjugate to a diagonal matrix. In the case of U(3) this would look like

$$u^{-1}vu = \begin{Bmatrix} e^{i\theta_1} & 0 & 0 \\ 0 & e^{i\theta_2} & 0 \\ 0 & 0 & e^{i\theta_3} \end{Bmatrix} = b, \ u \in U(3)$$

the idea of Weyl is to invert this relation so as to write any element in U(N) as a factorization

$$v = ubu^{-1} \quad (3)$$

However, the decomposition (3) is not unique. For $v$'s where two or more eigenvalues coincide the equation

$$ubu^{-1} = wcw^{-1} = v$$

will have solutions where $u$ and $w$ differ non-trivially.

We therefore restrict the analysis to that part U(N)' of U(N) for which all eigenvalues are different, that is the open and dense space for which

$$\theta_i \neq \theta_j \mod 2\pi \quad \text{for all} \ i \neq j \quad (4)$$

This restriction will not matter for the validity of our Laplacian to problems where the polar decomposition is applicable. The restriction is related to the fact that the metric is zero when $\theta_i = \theta_j$ for some $i$ and $j$. Although the Laplacian will be singular for $\theta_i = \theta_j$ just like the polar coordinate Laplacian is singular for r = 0 it will still be applicable on all of U(N) and SU(N) exactly for those problems which have "radial" (toroidal) symmetry. This is so because the submanifold for which $\theta_i = \theta_j$ for two or more $i, j$ has a measure 0 in U(N) and SU(N). Thus the singularity will vanish when the Laplacian is used in integration.

So let us consider a polar decomposition

$$(U(N)/A) \times A' \to U(N)' \equiv U' \quad (5)$$

where A is the subgroup of unitary diagonal matrices (the maximal torus). A' is the corresponding restriction with no coinciding eigenvalues and U(N)/A is the class of unitary matrices which are right equivalent modulus diagonal matrices. Weyl calls the elements of $U(N)/A \equiv \overline{U}$ for verticals and the elements of A' for horizontals (see fig.1)

## III. TRANSFORMATION OF VECTOR FIELDS

The map between the two spaces in (5) is given in accordance with (3) by the conjugation

$$\kappa: (\overline{u}, a) \to uau^{-1} \quad (6)$$

where u is a representative of the class element $\overline{u}$.

The tangent space decomposes

$$T_{(\overline{u},a)}(\overline{U} \times A') = T_{\overline{u}}(\overline{U}) \oplus T_a A'$$

That is, a direct sum of horizontal and vertical parts. Therefore the differential of $\kappa$ can be seen as a map from the decomposed tangent space

$$(d\kappa)_{(\overline{u},a)}: \ T_{\overline{u}}\overline{U} \oplus T_a A' \to T_{uau^{-1}}U' \quad (7)$$

and so, since the differential is linear, we need only consider the action of $d\kappa$ on the two individual parts (X,0) and (0, Y) where X is an element of the horizontal space and Y is an element of the vertical space.

In general an element of the Lie algebra generates vector fields on the manifold. The vector fields act as left invariant directional derivatives on functions $\Psi$ on the manifold by

$$Z_u[\Psi] \equiv Z[\Psi](u) = (d\Psi)_u(Z) = \frac{d}{dt}\Psi(u\exp(tZ))\Big|_{t=0} \quad (8)$$

With $Z = d\kappa(0,Y)$ the chain rule gives for the vertical part

$$d\kappa(0,Y)[\Psi](\bar{e},a) = (d\kappa)_{(\bar{e},a)}(0,Y)[\Psi]$$
$$= (d\Psi)_{\kappa(\bar{e},a)}((d\kappa)_{(\bar{e},a)}(0,Y))$$
$$= \frac{d}{dt}\Psi \circ \kappa(\bar{e}, a\exp tY)|_{t=0}$$
$$= \frac{d}{dt}\Psi(ea\exp tY e^{-1})|_{t=0}$$
$$= \frac{d}{dt}\Psi(a\exp tY)|_{t=0}$$
$$= (d\Psi)_a(Y) = \tilde{Y}[\Psi](a) \equiv \tilde{Y}_a[\Psi]$$

(9)

where $\tilde{Y}$ is the left invariant vector field generated by $Y$. Thus $d\kappa$ acts as the identity on the vertical vector fields. That is to say, $d\kappa$ acts as the identity on toroidal derivatives.

Now for the horizontal part

$$d\kappa(X,0)[\Psi](\bar{e},a)$$
$$= (d\kappa)_{(\bar{e},a)}(X,0)[\Psi]$$
$$= \frac{d}{dt}\Psi \circ \kappa(\overline{e\exp tX}, a)|_{t=0}$$
$$= \frac{d}{dt}\Psi((\exp tX)a(\exp tX)^{-1})|_{t=0}$$
$$= \frac{d}{dt}\Psi(a\exp(ta^{-1}Xa)\exp(-tX))|_{t=0}$$
$$= \frac{d}{dt}\Psi(a(I+ta^{-1}Xa+\cdots)(I-tX+\cdots))|_{t=0}$$
$$= \frac{d}{dt}\Psi(a(I+ta^{-1}Xa-tX+O(t^2)))|_{t=0}$$
$$= a(a^{-1}\tilde{X}a - \tilde{X})[\Psi] \equiv (a^{-1}\tilde{X}a - \tilde{X})_a[\Psi]$$

(10)

Thus $d\kappa$ acts in a more complicated way on the horizontal vector fields. We see that the resulting vector field is determined by the commutator

$$a(a^{-1}\tilde{X}a - \tilde{X}) = [\tilde{X},a]$$

Geometrically the Lie derivative $[X,a]$ measures the variation of $a$ along the dynamical system belonging to $X$. We may think of this as a generalized curl. The appearance of the commutator may not be so surprising since $\kappa$ maps by conjugation, i.e. a unitary rotation. And the differential of a unitary rotation is thus a commutator. We see an analogy with the generator of ordinary infinitesimal rotations about the z-axis in $R^3$ being[13]

$$L_z = y\frac{\partial}{\partial x} - x\frac{\partial}{\partial y}$$

which has a "curly" structure.

## IV. CHOICE OF BASIC DERIVATIVES

We are now ready to choose the basis of the two vector spaces to be used as derivatives in the expression for the Laplacian.

In general the algebras of U(N) and SU(N) consist of skew-hermitian matrices

$$Z^+ = -Z \qquad (11)$$

where the dagger indicates transposing and complex conjugation. The tangent space $T_a A'$ is simply given in the case of U(3) by

$$T_a A' = \{Z | Z = i\begin{pmatrix} e^{i\theta_1} & 0 & 0 \\ 0 & e^{i\theta_2} & 0 \\ 0 & 0 & e^{i\theta_3} \end{pmatrix} | \theta_i \in R\} \equiv \{iH(\boldsymbol{\theta})\}$$

and is mapped onto the torus as usual by the exponential

$$\exp: iH(\boldsymbol{\theta}) \to e^{iH(\boldsymbol{\theta})}$$

Thus the abelian vertical algebra (the socalled Cartan algebra) for instance of U(3) is generated by

$$T_1 = \begin{pmatrix} 1 & 0 & 0 \\ 0 & 0 & 0 \\ 0 & 0 & 0 \end{pmatrix}, T_2 = \begin{pmatrix} 0 & 0 & 0 \\ 0 & 1 & 0 \\ 0 & 0 & 0 \end{pmatrix} \text{ and } T_3 = \begin{pmatrix} 0 & 0 & 0 \\ 0 & 0 & 0 \\ 0 & 0 & 1 \end{pmatrix}$$

To construct a basis of the non-abelian horizontal algebra we start by the Lie algebra of gl(N,R) and choose a basis[14] such that

$$[A_{ij}, A_{kl}] = \delta_{jk}A_{il} - \delta_{li}A_{kj} \qquad (12)$$

and

$$A_{ji} = A_{ij}^T \tag{13}$$

That is, the non-commuting base elements come in transposed pairs.

This basis, which is proportional to the Cartan-Weyl basis (see the last section), can be represented by the matrices $E_{ij}$ which have zero elements except of the element 1 in row number i and coloumn number j, for example for u(3)

$$E_{32} = \begin{pmatrix} 0 & 0 & 0 \\ 0 & 0 & 0 \\ 0 & 1 & 0 \end{pmatrix}$$

Then

$$[H(\boldsymbol{\theta}), E_{ij}] = (\theta_i - \theta_j) E_{ij} \tag{14}$$

We see that only two toroidal parameters are involved. However $E_{ij}$ does not lie in the algebra of U(N) whereas the skew-hermitizised $E_{ij} - E_{ji}$ and $i(E_{ij} + E_{ji})$ do. We see explicitly that

$$E_{ij} - E_{ji} = E_{ij} - E_{ij}^+ = -(E_{ij} - E_{ij}^+)^+$$

and

$$i(E_{ij} + E_{ji}) = i(E_{ij} + E_{ij}^+) = -(i(E_{ij} + E_{ij}^+))^+ \tag{15}$$

show the skew-hermiticity.

We have

$$a = e^{iH(\boldsymbol{\theta})}$$

thus like in (14) we get

$$a^{-1} E_{ij} a = e^{-i\theta_i + i\theta_j} E_{ij}$$

Now return to the action of $d\kappa$ in (10). With $X = E_{ij} - E_{ji}$ we get

$$a^{-1}(E_{ij} - E_{ji})a - (E_{ij} - E_{ji})$$
$$= e^{-i(\theta_i - \theta_j)} E_{ij} - e^{-i(\theta_j - \theta_i)} E_{ji} - E_{ij} + E_{ji}$$
$$= (\cos(\theta_i - \theta_j) - 1)(E_{ij} - E_{ji}) - \sin(\theta_i - \theta_j) i(E_{ij} + E_{ji}) \tag{16}$$

And with $X = i(E_{ij} + E_{ji})$ we get

$$ia^{-1}(E_{ij} + E_{ji})a - i(E_{ij} + E_{ji})$$
$$= (\cos(\theta_i - \theta_j) - 1)i(E_{ij} + E_{ji}) + \sin(\theta_i - \theta_j)(E_{ij} - E_{ji}) \tag{17}$$

We see that with the choice (15), the horizontal basis vectors (only) mix pairwise. Note that the calculations were linear in $E_{ij}$ and $E_{ji}$, so that proper normalization introduces no extra factors.

**Normalization**

Talking about normalization it is now time to explicate the canonical metric

$$g(V, W) = -\mathrm{Tr}VW = \mathrm{Tr}VW^+ \tag{18}$$

where V and W belong to the Lie algebra u(N) of U(N).

The $N^2$ base vectors

$$iT_j, \tfrac{1}{\sqrt{2}}(E_{ij} - E_{ji}), \tfrac{i}{\sqrt{2}}(E_{ij} + E_{ji})$$
$$i \neq j, \ i, j = 1, 2, ..., N \tag{19}$$

are orthonormal with respect to this metric and they each represent a directional derivative $\partial_k$ in the Riemannian expression eq. (1) for the Laplacian. They also generate the vector fields $\tilde{X}_k$ and $\tilde{Y}_m$ in the Casimir representation in eq. (2) of the Laplacian on $U'$.

To simplify notation put

$$X_k = \tfrac{1}{\sqrt{2}}(E_{ij} - E_{ji}), \ \tilde{Z}_k(a) = (a^{-1}X_k a - X_k)_a$$
$$X_l = \tfrac{i}{\sqrt{2}}(E_{ij} + E_{ji}), \ \tilde{Z}_l(a) = (a^{-1}X_l a - X_l)_a$$

and

$$H_j = T_j \leftrightarrow \frac{\partial}{\partial \theta_j}$$

### V. METRIC COMPONENTS

Now we calculate the metric components for the Laplacian on U(N)' from the basis on $\bar{U} \times A'$. Because of the left invariance of the relevant vector fields, we may restrict ourselves to the case $\bar{u} = \bar{e}$ (see shortly).

As already mentioned in (7) $d\kappa$ maps from one tangent space to the other, and $\kappa$ maps by conjugation. This conjugation "mimics" the transformation of the coordinate fields $\tilde{Z}_k$ and $\tilde{Z}_l$ on all of U(N)' from their values on the torus

$$\tilde{Z}(v) = u\tilde{Z}(a)u^{-1} \tag{20}$$

where $v = uau^{-1}$.

With the transformation (20) we can show that the metric components are independent of the action of conjugation, i.e. they are constant along horizontals

$$\begin{aligned}g_{kl}(v) &= g(\tilde{Z}_k(v),\tilde{Z}_l(v)) \\ &= g(u\tilde{Z}_k(a)u^{-1}, u\tilde{Z}_l(a)u^{-1}) \\ &= -\text{Tr}(u\tilde{Z}_k(a)u^{-1}u\tilde{Z}_l(a)u^{-1}) \\ &= -\text{Tr}(\tilde{Z}_k(a)\tilde{Z}_l(a)) = g_{kl}(a)\end{aligned} \tag{21}$$

Thus, using (7) and inserting (16) and (17) in the above we get

$$\begin{aligned}g_{kl}(v) &= g_{kl}(a) \\ &= g(d\kappa_{(\bar{e},a)}(X_k(\bar{e}),0), d\kappa_{(\bar{e},a)}(X_l(\bar{e}),0)) \\ &= -\text{Tr}(d\kappa_{(\bar{e},a)}(X_k,0) d\kappa_{(\bar{e},a)}(X_l,0)) \\ &= \text{Tr}(((\cos(\theta_i - \theta_j)-1)X_k - \sin(\theta_i - \theta_j)X_l) \\ &\quad (\sin(\theta_i - \theta_j)X_k^+ + (\cos(\theta_i - \theta_j)-1)X_l^+)) \\ &= 0\end{aligned} \tag{22}$$

Similarly

$$\begin{aligned}g_{kk}(v) &= \text{Tr}(((\cos(\theta_i - \theta_j)-1)X_k - \sin(\theta_i - \theta_j)X_l) \\ &\quad ((\cos(\theta_i - \theta_j)-1)X_k^+ - \sin(\theta_i - \theta_j)X_l^+)) \\ &= 2(1-\cos(\theta_i - \theta_j)) \\ &= 4\sin^2\tfrac{1}{2}(\theta_i - \theta_j) = g_{ll}(v)\end{aligned} \tag{23}$$

Here the orthonormality of $X_k$ and $X_l$ has been used. For the vertical fields $\tilde{Y}_m(a) = aY_m = aiT_m$ the metrical components $g_{mm}$ are trivially equal to 1. So by the choice of the base vectors (19) we have succeeded to construct a diagonal polar metric on U(N)'. From (23) we readily get for the square root of the absolute value of the determinant

$$\sqrt{\|g(v)\|} = \prod_{i<j}^N 4\sin^2\tfrac{1}{2}(\theta_i - \theta_j) \equiv J^2 \tag{24}$$

where $J$ is the socalled van der Monde's determinant, also sometimes referred to as the Jacobian. For U(3) it is

$$\begin{aligned}J &= \prod_{i<j}^3 2\sin\tfrac{1}{2}(\theta_i - \theta_j) \\ &= -8\sin\tfrac{1}{2}(\theta_1 - \theta_2)\sin\tfrac{1}{2}(\theta_2 - \theta_3)\sin\tfrac{1}{2}(\theta_3 - \theta_1)\end{aligned} \tag{25}$$

Since g is diagonal the inverse $g^{-1}$ is diagonal too with diagonal components being simply the reciprocals from g

$$g^{kk} = g_{kk}^{-1} \tag{26}$$

## VI. THE GENERAL LAPLACIAN

We can now insert (19), (22)-(24) and (26) into (1) to get our final result

$$\begin{aligned}\Delta &= \sum_{j=1}^N \frac{1}{J^2}\frac{\partial}{\partial\theta_j} J^2 \frac{\partial}{\partial\theta_j} \\ &+ \sum_{i<j}^N \frac{\frac{1}{\sqrt{2}}(E_{ij}-E_{ji})\frac{1}{\sqrt{2}}(E_{ij}-E_{ji}) + \frac{i}{\sqrt{2}}(E_{ij}+E_{ji})\frac{i}{\sqrt{2}}(E_{ij}+E_{ji})}{4\sin^2\tfrac{1}{2}(\theta_i - \theta_j)}\end{aligned}$$

$$\tag{27}$$

Note that $E_{ij} - E_{ji}$ and $i(E_{ij} + E_{ji})$ belong to $T_{\bar{e}}\overline{U}$. In other words they act on a "constant" so-called representation space. They commute with the $\theta$-dependent metric components. A fact that showed up already in (21). We have thus succeeded to decompose the vertical and the horizontal derivatives and get a polarly decomposed Laplacian with a radial and an angular part.

Our result (27) has been derived using real matrices. In stead one could use a set defined by

$$iW_{ij} = E_{ij} \tag{28}$$

This would not change (14) and therefore would not change the structure of (27). The only change is in that the commutation relations of the operators in the numerator change to those of Wadia[12]. Rewriting the numerator of the second term

$$-\tfrac{1}{2}(E_{ij}E_{ji} + E_{ji}E_{ij}) - \tfrac{1}{2}(E_{ij}E_{ji} + E_{ji}E_{ij})$$
$$= -E_{ij}E_{ji} - E_{ji}E_{ij} = -E_{ij}E_{ij}^+ - E_{ji}E_{ji}^+$$

and inserting (28) we namely get

$$\Delta = \sum_{j=1}^N \frac{1}{J^2}\frac{\partial}{\partial\theta_j} J^2 \frac{\partial}{\partial\theta_j} - \sum_{i<j} \frac{E_{ij}E_{ij}^+ + E_{ji}E_{ji}^+}{4\sin^2\tfrac{1}{2}(\theta_i - \theta_j)}$$

$$= \sum_{j=1}^N \frac{1}{J^2}\frac{\partial}{\partial\theta_j} J^2 \frac{\partial}{\partial\theta_j} - \sum_{i\neq j} \frac{iW_{ij}(iW_{ij})^+}{4\sin^2\tfrac{1}{2}(\theta_i - \theta_j)}$$

$$= \sum_{j=1}^N \frac{1}{J^2}\frac{\partial}{\partial\theta_j} J^2 \frac{\partial}{\partial\theta_j} - \sum_{i\neq j} \frac{W_{ij}^2}{4\sin^2\tfrac{1}{2}(\theta_i - \theta_j)}$$

$$W_{ij}^2 \equiv W_{ij}W_{ij}^+$$
$$[W_{ij}, W_{kl}] = -i(\delta_{jk}W_{il} - \delta_{li}W_{kj})$$

(29)

We have now put our Laplacian in a form close to that of Wadia[12], with our $W_{ij}$ commuting as his $L_{ij}$. Wadia's radial part agrees with our result up to an additive constant, see eq. (33). He interpretes his $L_{ij}$ in the nominator of the angular part as off diagonal components of body fixed angular momentum. We have not been able to derive this interpretation since the $W_{ij}$s are not hermitian and therefore cannot represent physical observables. On the other hand our $E_{ij}$, being real an in transposed pairs, *are* hermitian, but their commutation relations have the opposite sign of Wadia's $L_{ij}$. (In the section on physical notation we will rewrite our general result into an expression including all the components of body fixed angular momentum in the specific cases of U(3) and SU(3)).

Finally we want to remind you that neither $E_{ij}$ nor $W_{ij}$ separately belong to the Lie algebra of U(N), whereas the operators (19) used in (27) do.

**Rewritten Laplacian**

For some practical purposes it is convenient to rewrite the radial part of (27) by

$$\sum_j \frac{1}{J^2}\frac{\partial}{\partial\theta_j} J^2 \frac{\partial}{\partial\theta_j} = \sum_j \left(\frac{1}{J}\frac{\partial^2}{\partial\theta_j^2} J - \frac{1}{J}\frac{\partial^2 J}{\partial\theta_j^2}\right) \quad (30)$$

For *U(3)* we can show by explicit differentiation of (25) that the last part on the right hand side is a constant

$$\sum_{j=1}^3 \frac{1}{J}\frac{\partial^2 J}{\partial\theta_j^2} = -2 \qquad (31)$$

The differentiations to prove this are greatly simplified by noting the identity

$$-4\sin\tfrac{1}{2}(x-y)\sin\tfrac{1}{2}(y-z)\sin\tfrac{1}{2}(z-x)$$
$$= \sin(x-y) + \sin(y-z) + \sin(z-x)$$

which is easily proven by using
$\sin\alpha = (e^{i\alpha} - e^{-i\alpha})/2i$.

The constant in (31) in general for *U(N)* is given by[1]

$$\sum_{j=1}^N \frac{1}{J}\frac{\partial^2 J}{\partial\theta_j^2} = -R_N = -\tfrac{1}{12}N(N^2-1) \qquad (32)$$

and is one sixth of the scalar curvature of the group manifold[15].

With the rewriting in (30) we then have the full Laplacian in a more practical and "physical" edition

$$\Delta = \sum_{j=1}^N \frac{1}{J}\frac{\partial}{\partial\theta_j^2} J + \frac{1}{12}N(N^2-1) - \sum_{i<j} \frac{L_{ij}^2 + M_{ij}^2}{8\sin^2\tfrac{1}{2}(\theta_i - \theta_j)}$$

$$iL_{ij} = E_{ij} - E_{ji}$$
$$iM_{ij} = i(E_{ij} + E_{ji})$$

(33)

where $L_{ij}$ and $M_{ij}$ are hermitian off-diagonal generators.

This edition of the Laplacian is analogous in structure to the one suggested by Wadia[12]. Wadia gave it up to a constant which we now recognize as an additive constant. From the interpretation below for the case of U(3) and SU(3) we may think of the $L_{ij}$s as components of generalized body fixed angular momentum operators. Note that $L_{ij}$ and $M_{ij}$ do not have simple commutation relations like (12) and (29) for $E_{ij}$ and $W_{ij}$.

## The Laplacian on SU(N)

The Laplacian (27) or (33) may be used also for SU(N) with the restriction that

$$\theta_1 + \cdots + \theta_N \equiv 0 \mod 2\pi \qquad (34)$$

One may worry about the redundancy of parametrizing the N-1-dimensional torus of SU(N) with N angular parameters, but as long as the constraint (34) is maintained we are in the same legitimate situation as when the rectangular coordinate Laplacian

$$\Delta = \frac{\partial^2}{\partial x^2} + \frac{\partial^2}{\partial y^2} + \frac{\partial^2}{\partial z^2}$$

is used to formulate and solve problems on the surface of a sphere under the constraint that

$$x^2 + y^2 + z^2 = r^2$$

## VII. THE LAPLACIAN IN PHYSICAL NOTATION

We have already introduced a notation in equation (33) which accords with the physical convention of expressing generators as hermitian operators times the imaginary unit. For convenience we will here introduce an even more simplified notation for U(3) and SU(3). We introduce

$$iL_3 = E_{12} - E_{21} = \begin{pmatrix} 0 & 1 & 0 \\ -1 & 0 & 0 \\ 0 & 0 & 0 \end{pmatrix} = i\lambda_2$$

$$iL_2 = E_{13} - E_{31} = i\lambda_5$$
$$iL_1 = E_{23} - E_{32} = i\lambda_7$$
$$iM_3 = i(E_{12} + E_{21}) = i\lambda_1$$
$$iM_2 = i(E_{13} + E_{31}) = i\lambda_4$$
$$iM_1 = i(E_{23} + E_{32}) = i\lambda_6$$

(35)

Here we recognize $L_k$ and $M_k$ as the off-diagonal, well known Gell-Mann matrices $\lambda^{16}$.

The reason we want in physics to express the operators as hermitian is that we may want them to serve as *observables* which implies that they must have real eigenvalues. A characteristic which is met by requiring hermiticity.

Thus, using the skew-hermiticity implied by (15) we can rewrite our result (27) into a more compact form

$$\Delta = \sum_{j=1}^{3} \frac{1}{J^2} \frac{\partial}{\partial \theta_j} J^2 \frac{\partial}{\partial \theta_j} - \sum_{\substack{i<j \\ k \neq i,j}}^{3} \frac{(iL_k)(iL_k)^+ + (iM_k)(iM_k)^+}{8 \sin^2 \tfrac{1}{2}(\theta_i - \theta_j)}$$

or

$$\Delta = \sum_{j=1}^{3} \frac{1}{J^2} \frac{\partial}{\partial \theta_j} J^2 \frac{\partial}{\partial \theta_j} - \sum_{\substack{i<j \\ k \neq i,j}}^{3} \frac{L_k^2 + M_k^2}{8 \sin^2 \tfrac{1}{2}(\theta_i - \theta_j)}$$

(36)

where $L_k^2 \equiv L_k L_k^+$ with no summation over k and similarly for $M_k$. (36) shows a very close structural similarity to the well known polar coordinate Laplacian with which the present goal was compared.

The components of

$$\mathbf{L} = (L_1, L_2, L_3) \qquad (37)$$

commute as body fixed angular momentum operators

$$[L_k, L_l] = -iL_m \qquad (k,l,m \text{ cyclic}) \qquad (38)$$

Thus the structure of (36) reflects a toroidal part generated by the abelian algebra $T_j$, an angular momentum part generated by an *su(2)* subalgebra, and a mixing part generated by the remaining three generators $M_k$ which commute "out of their own subspace"

$$[M_k, M_l] = -iL_m \qquad (39)$$

They so to say glue the total algebra together and thereby express the simple nature of *su(3)*.

Note that $L_k$ and $M_k$ commute individually with the radial part. This can be shown directly from the commutation relations. $L_k$ and $M_k$ obviously commute with $J$ and for instance for $L_1$ we have $[L_1, T_1^2] = 0$ and $[L_1, T_2^2] = iM_1$ is cancelled by $[L_1, T_3^2] = -iM_1$. It may come as a bigger surprise that $L_k$ and $M_k$ also commute with the angular part. For instance $[L_1, M_1^2] = 0$ and the commutator of $L_1$ with the $L_2^2 + M_2^2$ term will vanish since the $L_2^2$ and $M_2^2$ terms have the same denominator and since $[L_1, L_2^2] = iM_1 = -[L_1, M_2^2]$. Actually these specific facts are no surprise since they only reflect the general criterion that the Laplacian, being a Casimir operator, should commute with all the generators[17]. As $L_k$ and $M_k$ namely must commute with the total Laplacian and are seen to commute with the abelian/toroidal/radial part, it follows that they must commute with the angular part too. It is thus clear that the general criterion is met separately by the radial and angular parts.

For practical purposes it is worth to notice that

$$[L_k, \mathbf{L}^2] = 0 \qquad (40)$$

as allready known from the *su(2)* algebra. It can be shown also that

$$[L_k, \mathbf{M}^2] = [M_k, \mathbf{M}^2] = [M_k, \mathbf{L}^2] = 0 \qquad (41)$$

Using (30) and (31) for U(3) and SU(3) we get a specific edition of (33)

$$\Delta = \sum_{j=1}^{3} \frac{1}{J} \frac{\partial^2}{\partial \theta_j^2} J + 2 - \sum_{\substack{i<j \\ k \neq i,j}} \frac{L_k^2 + M_k^2}{8\sin^2 \frac{1}{2}(\theta_i - \theta_j)} \qquad (33')$$

**Full basis and commutators**

For your convenience we give here the total basis in the representation chosen

$$T_1 = \begin{pmatrix} 1 & 0 & 0 \\ 0 & 0 & 0 \\ 0 & 0 & 0 \end{pmatrix}, \quad T_2 = \begin{pmatrix} 0 & 0 & 0 \\ 0 & 1 & 0 \\ 0 & 0 & 0 \end{pmatrix}, \quad T_3 = \begin{pmatrix} 0 & 0 & 0 \\ 0 & 0 & 0 \\ 0 & 0 & 1 \end{pmatrix}$$

$$L_1 = \begin{pmatrix} 0 & 0 & 0 \\ 0 & 0 & -i \\ 0 & i & 0 \end{pmatrix}, \quad L_2 = \begin{pmatrix} 0 & 0 & -i \\ 0 & 0 & 0 \\ i & 0 & 0 \end{pmatrix}, \quad L_3 = \begin{pmatrix} 0 & -i & 0 \\ i & 0 & 0 \\ 0 & 0 & 0 \end{pmatrix}$$

$$M_1 = \begin{pmatrix} 0 & 0 & 0 \\ 0 & 0 & 1 \\ 0 & 1 & 0 \end{pmatrix}, \quad M_2 = \begin{pmatrix} 0 & 0 & 1 \\ 0 & 0 & 0 \\ 1 & 0 & 0 \end{pmatrix}, \quad M_3 = \begin{pmatrix} 0 & 1 & 0 \\ 1 & 0 & 0 \\ 0 & 0 & 0 \end{pmatrix}$$

The mnemotecnic advantages of our proposed notation should be clear. For instance $L_3$ and $M_3$ generate unitary rotations around the 3-"axis".

We give also the list of commutators of the angular momemtum and mixing operators among each other and with the toroidal operators to supplement (38) and (39). Commutators not listed are zero

$[L_1, M_2] = [L_2, M_1] = -iM_3$
$[L_1, M_3] = -[L_3, M_1] = iM_2$
$[L_2, M_3] = [L_3, M_2] = iM_1$

$[L_1, M_1] = 2i(T_3 - T_2)$
$[L_2, M_2] = 2i(T_3 - T_1)$
$[L_3, M_3] = 2i(T_2 - T_1)$

$[L_1, T_2] = iM_1; [L_1, T_3] = -iM_1$
$[L_2, T_3] = -iM_2; [L_2, T_1] = iM_2$
$[L_3, T_1] = iM_3; [L_3, T_2] = -iM_3$

$[M_1, T_2] = -iL_1; [M_1, T_3] = iL_1$
$[M_2, T_3] = iL_2; [M_2, T_1] = -iL_2$
$[M_3, T_1] = -iL_3; [M_3, T_2] = iL_3$

This crazy list together with (38) and (39) is of course contained in the structure constants of *U(3)*, but you may enjoy to glance at the list in its totality. It gives a sense of the inherent triplicity in the algebraic structure which is here made explicit by the two class labels $L$ and $M$ and their common index $k$.

**The angular part in other well known basisses**

The Gell-Mann matrices $\lambda_i$ may be defined by linear combinations from the so-called Cartan-Weyl basis[18] whose specific construction from roots and step operators is given by Fonda and Ghirardi[19].

The so defined generators are traditionally labelled $F_i = \tfrac{1}{2}\lambda_i$, $i = 1,...,8$ where $F_3 = \tfrac{1}{2}\lambda_3 = H_1$ and $F_8 = \tfrac{1}{2}\lambda_8 = H_2$ is a maximal set of two commuting elements of the rank two algebra of SU(3). For the off-toroidal generators we have from (35) the identifications

$$
\begin{array}{ll}
L_3 = 2F_2 & M_3 = 2F_1 \\
L_2 = 2F_5 & M_2 = 2F_4 \\
L_1 = 2F_7 & M_1 = 2F_6
\end{array}
\qquad (42)
$$

It is thus straightforward to substitute these into the angular terms of the Laplacian to get for instance

$$\frac{L_3^2 + M_3^2}{8\sin^2 \tfrac{1}{2}(\theta_1 - \theta_2)} = \frac{F_2^2 + F_1^2}{2\sin^2 \tfrac{1}{2}(\theta_1 - \theta_2)}$$

The non-diagonal elements $E_\alpha$ of the Cartan-Weyl basis are defined by

$$[H_i, E_\alpha] = \alpha_i E_\alpha \qquad (43)$$

where $\alpha_i$ are the roots[20]. The $F_i$s are defined simply as a change of basis in the algebra to give simple commutation relations

$$[F_i, F_j] = \sum_k f_{ijk} F_k \qquad (44)$$

where $f_{ijk}$ are the structure constants of SU(3). Following Fonda and Ghirardi we thus have

$$
\begin{aligned}
M_3 &= \sqrt{6}(E_{\alpha^{(1)}+\alpha^{(2)}} + E_{-(\alpha^{(1)}+\alpha^{(2)})}) \\
L_3 &= -i\sqrt{6}(E_{\alpha^{(1)}+\alpha^{(2)}} - E_{-(\alpha^{(1)}+\alpha^{(2)})}) \\
M_2 &= \sqrt{6}(E_{\alpha^{(1)}} + E_{-\alpha^{(1)}}) \\
L_2 &= -i\sqrt{6}(E_{\alpha^{(1)}} - E_{-\alpha^{(1)}}) \\
M_1 &= \sqrt{6}(E_{\alpha^{(2)}} + E_{-\alpha^{(2)}}) \\
L_1 &= -i\sqrt{6}(E_{\alpha^{(2)}} - E_{-\alpha^{(2)}})
\end{aligned}
\qquad (45)
$$

or

$$
\begin{aligned}
E_{\alpha^{(1)}+\alpha^{(2)}} &= \tfrac{1}{2\sqrt{6}}(M_3 + iL_3) = \tfrac{1}{\sqrt{6}} I_+ \\
E_{-(\alpha^{(1)}+\alpha^{(2)})} &= \tfrac{1}{2\sqrt{6}}(M_3 - iL_3) = \tfrac{1}{\sqrt{6}} I_- \\
E_{\alpha^{(1)}} &= \tfrac{1}{2\sqrt{6}}(M_2 + iL_2) = \tfrac{1}{\sqrt{6}} V_+ \\
E_{-\alpha^{(1)}} &= \tfrac{1}{2\sqrt{6}}(M_2 - iL_2) = \tfrac{1}{\sqrt{6}} V_- \\
E_{\alpha^{(2)}} &= \tfrac{1}{2\sqrt{6}}(M_1 - iL_1) = \tfrac{1}{\sqrt{6}} U_- \\
E_{-\alpha^{(2)}} &= \tfrac{1}{2\sqrt{6}}(M_1 + iL_1) = \tfrac{1}{\sqrt{6}} U_+
\end{aligned}
\qquad (46)
$$

where the raising operators are associated with the positive roots[21] which for SU(3) are

$$\alpha^{(1)} = (\tfrac{1}{2}, \tfrac{\sqrt{3}}{2}),\ \alpha^{(2)} = (\tfrac{1}{2}, -\tfrac{\sqrt{3}}{2}),\ \alpha^{(1)} + \alpha^{(2)} = (1,0) \text{ and}$$

the letters $I$, $U$ and $V$ refer to the socalled $I$, $U$ and $V$-spin. Note that $U_-$ is a raising operator. Inverting (35) and comparing with (46) we see that $I$, $U$ and $V$ is just another notation for the basis we used to express our original Laplacian (27).

$$
\begin{aligned}
I_+ &= \tfrac{1}{2}(M_3 + iL_3) = E_{12} \\
I_- &= \tfrac{1}{2}(M_3 - iL_3) = E_{21} \\
V_+ &= \tfrac{1}{2}(M_2 + iL_2) = E_{13} \\
V_- &= \tfrac{1}{2}(M_2 - iL_2) = E_{31} \\
U_+ &= \tfrac{1}{2}(M_1 + iL_1) = E_{23} \\
U_- &= \tfrac{1}{2}(M_1 - iL_1) = E_{32}
\end{aligned}
\qquad (47)
$$

Thus $L_k$ and $M_k$ can be expressed as combinations of raising and lowering operators.

$$
\begin{array}{ll}
iL_3 = I_+ - I_- & iM_3 = i(I_+ + I_-) \\
iL_2 = V_+ - V_- & iM_2 = i(V_+ + V_-) \\
iL_1 = U_+ - U_- & iM_1 = i(U_+ + U_-)
\end{array}
\qquad (48)
$$

In some applications it might be desirable to express the angular part of the Laplacian in the notation of the *I*, *U* and *V* spin operators. For the *I* spin[22]

$$I_1 = \tfrac{1}{2}(I_+ + I_-) = \tfrac{1}{2} M_3$$
$$I_2 = \tfrac{1}{2i}(I_+ - I_-) = \tfrac{1}{2} L_3 \quad (49)$$
$$I_3 = F_3$$

and we have

$$I^2 = I_1^2 + I_2^2 + I_3^2 = \tfrac{1}{4}(L_3^2 + M_3^2) + I_3^2$$

Thus

$$\frac{L_3^2 + M_3^2}{8\sin^2 \tfrac{1}{2}(\theta_1 - \theta_2)} = \frac{I^2 - I_3^2}{2\sin^2 \tfrac{1}{2}(\theta_1 - \theta_2)}$$

Apart from giving a convenient overview of different basisses the point of this last section has been to show you that once the angular part has been derived in a basis natural for the derivation, you are free to choose a basis suitable for your specific purpose.

## VIII. CONCLUDING REMARKS

We have derived an "operational" expression for the full Laplacian on U(N) based on elementary differential geometry and using polar decomposition. We have suggested simplifying notations for the Laplacian on SU(3) and U(3).

## ACKNOWLEDGMENTS

The first name author would like to thank Karen Ter-Martirosyan for putting him in contact with Yurii Makeenko, Nikita Nekrasov and Dmitri Boulatov during "the Russians autumn stay" at the Niels Bohr Institute, Denmark in November 1993. They stressed that one should expect an analogy with the polar coordinate Laplacian and gave an important reference to Barut and Raczka's marvellous book.

The same author would like to thank Adriano Di Giacomo for showing interest in the Laplacian at the conference on Quark Confinement and the Hadron Spectrum in Como, Italy 1994 and again in the final phase of preparing for publication.

## NOTES

The present paper was started as the last name author was still at Roskilde University, Denmark. His part has been mainly consulting. The first name author cannot agree on this modest description.

[12] S.R. Wadia, *N = ∞ Phase Transition in a Class of Exactly Soluble Model Lattice Gauge Theories*, Phys. Lett. **93B**, 403-410 (1980).
[13] B.L. van der Waerden, *Group Theory and Quantum Mechanics* (Springer, Berlin, Germany, 1974), p. 107.
[14] See Ref. 6, p. 254.
[15] See Ref. 3, p. 368.
[16] See Ref. 9, pp. 209.
[17] F.A. Berezin, *Laplace Operators on Semisimple Lie Groups*, Amer. Math. Soc. Transl. **21**(2), 239-339 (1962).
[18] See Ref. 6, p. 253.
[19] L. Fonda and G.C. Ghirardi, *Symmetry Principles in Quantum Physics* (Marcel Dekker, New York, USA, 1970), pp. 169, 181, 192.
[20] See Ref. 19, pp. 171, 193.
[21] See also Ref. 6, p. 255.
[22] See Ref. 19, pp. 191.

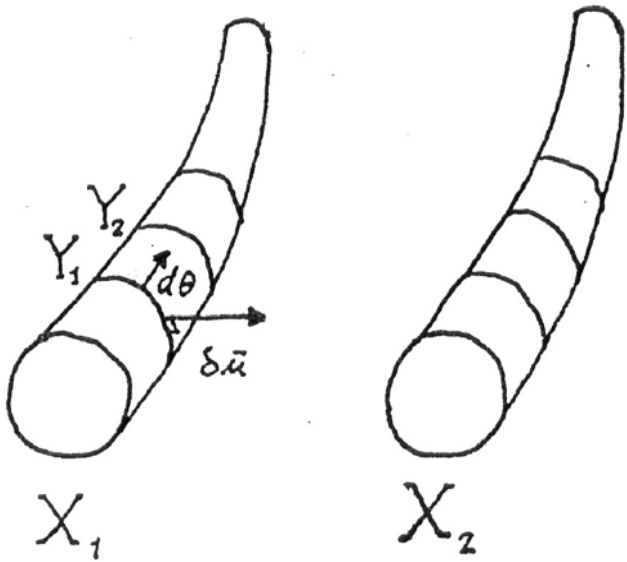

**Figure captions**

Figure 1: *Weyl's geometrical terminology. Think of toroidal increments as vertical (radial) increments and off-torus increments as horizontal (lateral) at right angles to the torus.*